\documentclass[sigconf]{acmart}

\AtBeginDocument{%
  \providecommand\BibTeX{{%
    \normalfont B\kern-0.5em{\scshape i\kern-0.25em b}\kern-0.8em\TeX}}}

\acmConference[WTF '23]{make sure it's right}{July 19,
  2023}{Eindhoven, The Netherlands}

\begin{document}

\title{Building for Speech: Designing the Next Generation of Social Robots for Audio Interaction}

\author{Angus Addlesee}
\email{a.addlesee@hw.ac.uk}
\affiliation{%
  \institution{Heriot-Watt University}
  \city{Edinburgh}
  \country{Scotland, UK}
}

\author{Ioannis Papaioannou}
\email{ioannis@alanaai.com}
\affiliation{%
  \institution{Alana AI}
  \city{Edinburgh}
  \country{Scotland, UK}
}

\author{Oliver Lemon}
\email{o.lemon@hw.ac.uk}
\affiliation{%
  \institution{Heriot-Watt University}
  \city{Edinburgh}
  \country{Scotland, UK}
}


\begin{abstract}
  There have been incredible advancements in robotics and spoken dialogue systems (SDSs) over the past few years, yet we still don't find social robots in public spaces like train stations, shopping malls, or hospital waiting rooms. In this paper, we argue that early-stage collaboration between robot designers and SDS researchers is crucial to create social robots that can legitimately be used in real-world environments. We draw from our experiences running experiments with social robots, and the surrounding literature, to highlight recurring issues. Robots need more speakers, more microphones, quieter motors, and quieter fans to enable human-robot spoken interaction in the wild and improve accessibility. More robust robot joints are also needed to limit potential harm to older adults and other more vulnerable groups.
\end{abstract}

%
%


%


\received{5 June 2023}
\received[accepted]{13 June 2023}
\received[revised]{5 July 2023}

\maketitle

\section{Introduction}

\begin{figure}
    \centering
    \includegraphics[width=0.47\textwidth]{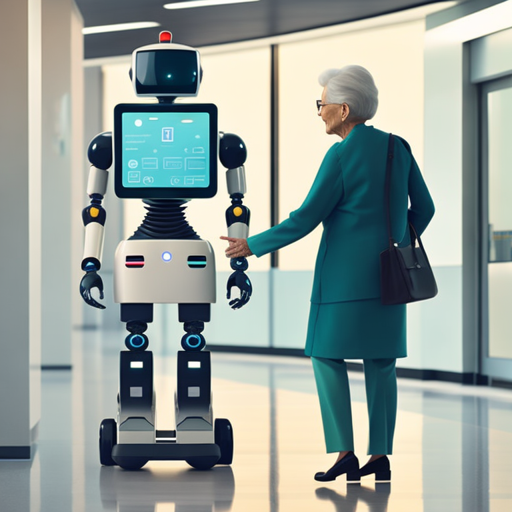}
    \caption{A person asking for directions in a hospital.}
    \label{fig:directions}
\end{figure}

Social robots are not yet found in our public spaces, despite this vision being an imminent reality over 25 years ago \cite{thrun1998robots}. They don't roam our shopping malls helping lost families find the bathrooms, we don't bump into them in museums and airports, and they are not helping patients in hospital waiting rooms (see Figure \ref{fig:directions}). 

Spoken dialogue systems (SDSs) have consistently improved over time \cite{glass1999challenges,williams2009spoken,lemon2022conversational}, with many years of peer-reviewed papers containing remarkable results. These models, however, are often evaluated automatically upon collected data, or with users in highly controlled lab settings. Social robots work wonderfully in the lab, but fail when deployed in the real world for experiments or demonstration. Some of these failures stem from the embedded SDS (for example: multi-party interactions \cite{addlesee2023data}, or voice accessibility \cite{addlesee2023voice}), but even interactions that the SDS should be able to handle with ease go wrong. These failures are often caused by the design of the social robot itself. The field of robotics has also seen incredible advancements over the past years, today's robots can navigate obstacle courses \cite{xiao2022autonomous}, manipulate objects in their environment \cite{chai2022survey}, generate human-like gestures \cite{tatarian2022does}, and follow complex human instructions using large language models (LLMs) \cite{ahn2022can}. Sadly, while collaboration between these two fields is common, they often begin after the robot has been designed. In our experience from multiple international robot dialogue projects, spoken interaction is not considered during the initial design phase. This lack of early-stage collaboration between robot designers and SDS researchers leaves room for oversight of critical features for spoken interaction. In this paper, we have combed the literature and drawn from our own experiences to highlight the main problems that repeatedly surface when experimenting with social robots and real users. We hope this paper sparks discussion between both communities to create fully functioning social robots that do genuinely work in public spaces in the future.

\section{People Struggle to Hear Robots}
\label{sec:speaker}

The first issue that crops up commonly in the literature is the limited volume of the robot's voice. Robot designers simply attach a speaker to the robot without considering the fact that the world is noisy, and some users, such as older adults, may have hearing loss.

In recent work, researchers deployed a robot to interact with real users in an assisted living facility. The robot had to be fitted with an additional speaker that had a louder maximum volume. The modification was necessary because the users simply could not hear the robot's voice, preventing any basic interaction \cite{stegner2023situated}.

This issue is not constrained to this particular setting, or to one particular robot. Researchers had to repeat every sentence the robot said in the lobby of a concert hall, as participants couldn't hear it \cite{langedijk2020studying}. In various school environments, the robot's volume was not loud enough to enable effective interaction, so external speakers had to be fitted \cite{nikolopoulos2011robotic}. When guiding people in an elder care facility, the robot's single speaker faced the wrong direction, so users could not hear it \cite{langedijk2020studying}. A robot was deployed in the homes of a few older adults, and they noted that its volume was not loud enough. People couldn't hear the social robot in a gym \cite{sackl2022social}, and the list goes on.

Robots are expensive. The speakers that researchers had to retrofit to the robots were inexpensive and readily available. This low-cost change was simple, yet \emph{crucial}, to enable effective communication with a user in a real-world setting. When designing robots for spoken interaction, we recommend fitting multiple speakers (facing various directions) that have a loud maximum volume. This will guarantee that the robot can be heard in public spaces, and ensure its accessibility for people with limited hearing.

\section{Robots Struggle to Hear People}
\label{sec:microphone}

There is another conversation participant that cannot hear what their interlocutor is saying -- the robot. This problem is similar to the one in Section \ref{sec:speaker}, and is also frequently found in the literature. A social robot struggled to hear users in a hotel lobby, for example \cite{hahkio2020service}. Many researchers retrofit better microphones to the robot \cite{villalpando2018predictive}, or next to the robot \cite{wagner2023comparing}, in order to hear the user more clearly.

In an assisted living facility, researchers had to resort to listening to the user through an ajar door to run their experiments. The microphone array could not reliably pick up what users said \cite{stegner2023situated}.

Home voice assistants do successfully hear people in noisy environments, like family homes \cite{porcheron2018voice}, however. They can pick up what the user said when other conversations are happening in the room, and when the TV or radio are on (we are finding this in ongoing work \cite{addlesee2022securely}). Today's social robots typically have four microphones\footnote{We checked the technical specifications of several commonly used social robots, and robots that we have deployed ourselves. We are refraining from naming specific robot creators, as this paper aims to encourage collaboration, and not criticise specific robots.}, but we argue that this is far too few. Apple's Homepod originally had six microphones \cite{calore2019review} and Amazon's Alexa Echo had seven \cite{spekking2021amazon}. The newest Homepod and Echo's have reduced to four microphones for two reasons: (1) These devices are incentivised to keep their device's costs low to encourage adoption by new users \cite{welch2023apple}; and (2) The device's shape and internal component arrangements have been refined and optimised over many years through experiments with millions of users \cite{wilson2020why}. Robots do not share either of these arguments. Microphones are trivially inexpensive relative to the price of a robot, and instead of helping microphones, the robot's shape actively hinders their performance. The body parts of a social robot often sit between the user and the microphone (for example, when the user is behind the robot, or in a wheelchair). Robots also create a lot of noise themselves, called \textit{ego-noise}. Related research required high-quality audio input from a noisy propellered UAV, so they attached sixteen microphones in various locations around the device \cite{nakadai2017development}, not just four.

Human-robot spoken communication can also be disrupted by societal or linguistic phenomena, such as overlapping or poorly formed turn-taking conditions. Such conditions include barging-in \cite{bargein} (where the user interrupts the robot mid-sentence, but the robot fails to recognise the user started speaking), and poor end-of-turn detection (due to long pauses or intermittent speech from the user). In our experience, users sometimes barge-in because of high latency caused by limited computational power onboard the robot, or on-site connectivity issues, in addition to the latency of the SDS.

\begin{figure}
    \centering
    \includegraphics[width=0.47\textwidth]{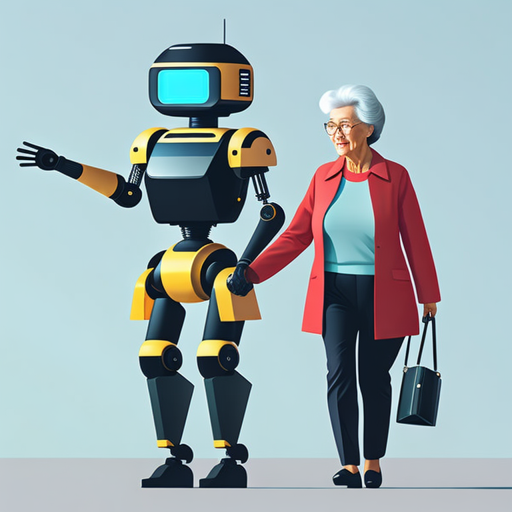}
    \caption{A social robot providing stability to an older adult.}
    \label{fig:helper}
\end{figure}

Potential approaches to this challenge include incremental dialogue processing \cite{addlesee2020comprehensive,aylett2023my}, or explicit turn-taking signals which enable the user to better understand when the robot is actually listening to them. For instance, \citet{foster2019mummer} employed a tablet on the robot's torso that was showing \textit{"I am listening"} and \textit{"I am speaking"} text to help guide the users in a noisy shopping mall setup.

We recommend fitting multiple high-quality microphones in various locations around the robot's body, as well as appropriate signal processing techniques, such as beam-forming \cite{adel2012beamforming}. Latency issues must be addressed within the SDS, and by increasing the robot's computational capabilities. These changes will again ensure that spoken interactions can realistically take place in public spaces. Designers must also consider microphone placement lower down on the robot for shorter users, and for people in wheelchairs, as they are commonly just placed on the top of the robot's head.

\section{Ego-Noise}
\label{sec:noisy}

This issue of ego-noise, introduced in Section \ref{sec:microphone}, is so problematic that an entire field of research has grown to tackle it. Researchers find that ego-noise doesn't just negatively impact ASR performance, but ego-noise reduction methods also suppress some of the user's utterance \cite{ince2010hybrid,schmidt2018novel}. To clarify, both the ego-noise reduction techniques, and the ego-noise itself negatively impact ASR performance.

This issue would be helped by additional speakers and microphones, ideally not placed next to noise sources, allowing both parties to hear each other. An optimal social robot designed for spoken interaction would also have much quieter joint motors and fans. These are more expensive than speakers and microphones, but they would greatly improve the SDSs ability to understand the user. This could be paired with research to repair and understand disrupted sentences \cite{addlesee2023understandingamr,addlesee2023understandingsparql}, while quieter motors are developed.

\section{Joint Robustness}

Ego noise obviously does not impact robots that do not have a body. In our minds, though, social robots should be able to point and guide users. For example, consider a hospital waiting room in a hospital memory clinic \cite{gunson2022visually}. Patients are typically older adults, and may use the robot's arm for stability, like they would with another human (see Figure \ref{fig:helper}). Current social robots can generate social gestures like waving or holding its hand out for a handshake. If you were to shake the robot's hand, however, it would likely break.

This fragility could potentially harm users if deployed in this setting. People may assume that they can link arms with the robot while being guided, a perfectly natural assumption. When an older adult puts their weight on the robot's joint, though, they would fall. This is clearly a potentially harmful design flaw that must be resolved if we are ever going to find robot assistants in the wild.

\section{Conclusion}

Interacting with social robots in public spaces is currently still a sci-fi fantasy. There are challenges that SDS researchers must tackle to reach this goal, but that is not the only bottleneck. Even a perfect SDS would fail if it was embedded in today's robots. We have highlighted that robot's need more speakers, more microphones, quieter fans, and quieter motors to allow both parties to hear each other. These are critical problems that completely block spoken interactions outside a lab setting. We additionally highlighted that robots also need to be more robust if they are to be safely applied in the real world, particularly in settings with older adults.

Our suggestions are not an exhaustive list, but we hope that they spark discussion and encourage collaboration between robot designers and SDS researchers. This collaboration should take place in the initial-stages of a robot's design to avoid the retrofitting of hardware discussed in this paper.

\begin{acks}
The images in this paper were generated with \url{Hotpot.io}.
\end{acks}

\bibliographystyle{ACM-Reference-Format}
\bibliography{sample-base}

\end{document}